\documentclass[fleqn,a4paper,prd,preprint,nofootinbib]{revtex4-1}

\usepackage{amsmath,amssymb}
\usepackage{longtable}
\usepackage[dvipdfmx]{graphicx,color}

\newcommand{\lagrangian}{\mathcal{L}}
\newcommand{\M}{{\mathcal M}}
\newcommand{\g}{{\mathfrak g}}
\newcommand{\h}{{\mathfrak h}}
\renewcommand{\S}[1]{\mathcal{S}_{#1}}

\newcommand{\Orb}[1]{\mathcal{O}_{#1}}
\newcommand{\C}[1]{{\mathcal C}_{#1}}
\newcommand{\bxi}{\boldsymbol{\xi}}
\newcommand{\bP}[1]{\boldsymbol{P}\kern-1pt_{#1}}
\newcommand{\bK}[1]{\boldsymbol{K}\kern-1pt_{#1}}
\newcommand{\bL}[1]{\boldsymbol{L}_{#1}}
\newcommand{\dS}{dS^2}
\newcommand{\AdS}{AdS^2}
\newcommand{\pf}[1]{{#1}_{\ast}}
\newcommand{\GL}[2]{\text{GL}(#1,#2)}
\newcommand{\R}{\mathbb{R}}
\newcommand{\Minkowski}{E^{3,1}}
\newcommand{\Vc}{{\cal V}_{\mathrm{C}}}
\newcommand{\Vs}{{\cal V}_{\mathrm{S}}}
\newcommand{\VIzero}{$\text{VI}_{\text{0}}$}
\newcommand{\VIIzero}{$\text{VII}_{\text{0}}$}
\newcommand{\VIIzeroI}{${\text{VII}_{\text{0}}}$--1}
\newcommand{\VIIzeroII}{${\text{VII}_{\text{0}}}$--2}
\newcommand{\VIII}{$\text{VIII}$}
\newcommand{\IX}{$\text{IX}$}
\newcommand{\pair}[2]{(#1,\,#2)}
\newcommand{\triple}[3]{(#1,\,#2,\,#3)}
\newcommand{\commutator}[2]{[#1,\,#2]}
\newcommand{\theorbits}{\VIIzero-orbits}
\newcommand{\ch}{\cosh\theta}
\newcommand{\sh}{\sinh\theta}
\newcommand{\co}{cohomogeneity one}

\DeclareMathOperator{\Isom}{Isom}
\newcommand{\wv}{world-volume}
\newcommand\inv{^{-1}}

\begin{document}
\title{Membranes with a symmetry of {\co}}

\author{Hiroshi Kozaki}
\email{kozaki@ishikawa-nct.ac.jp}
\affiliation{%
 Department of General Education, Ishikawa National College of
 Technology, Tsubata, Ishikawa 929-0392, Japan}

\author{Tatsuhiko Koike}
\email{koike@phys.keio.ac.jp}
\affiliation{%
 Department of Physics and RECNS, Keio University, Yokohama 223-8522, Japan}

\author{Hideki Ishihara}
\email{ishihara@sci.osaka-cu.ac.jp}
\affiliation{%
 Department of Mathematics and Physics,
 Graduate school of Science, Osaka City University,
 Osaka 558-8585, Japan}

\preprint{OCU-PHYS 412, AP-GR 116}

\begin{abstract}
  We study the dynamics of the Nambu-Goto membranes with  {\co}
  symmetry, 
  i.e., the membranes whose trajectories are foliated by homogeneous
  surfaces. 
  It is shown that the equation of motion reduces to a geodesic
  equation on a certain manifold, which is constructed from the
  original spacetime and Killing vector fields thereon. 
  A general method is presented for classifying 
  the symmetry of {\co} membranes 
  in a given spacetime. 
  The classification is completely carried out in Minkowski spacetime. 
  We analyze one of the obtained classes in depth and 
  derive an exact solution. 
\end{abstract}

\maketitle

\section{Introduction}
Extended objects come out in various areas of physics:
topological defects in field theories 
and in condensed matter physics, and branes in string theories, recently.
In cosmology, topological defects such as cosmic strings
and domain walls are supposed to have formed in the early universe.
In the brane-world universe models,
the universe itself is an extended object embedded in a bulk space
\cite{Randall:1999ee}. 
Recently, configuration of extended objects in black hole spacetimes
are also of growing importance in discussions of strong coupling
regime of gauge theories through gauge/gravity
duality~\cite{Maldacena:1997re}.

Extended objects, compared with particles,
have a wide variety of motion. 
For example, in Minkowski spacetime, a free particle moves with a
constant velocity so that its only possible trajectories are timelike
straight lines.
On the other hand, the trajectories of a string can 
be two-dimensional
timelike surfaces with various deformations. 
It is of fundamental importance to clarify the possible motion of
extended objects in a given spacetime.
However, we do not know much
because the equations of motion (EOM) are difficult to solve;
the EOM for extended objects are partial differential
equations (PDEs) 
while those for particles are ordinary differential equations
(ODEs). 
Even in the case of strings, 
where EOM are written as PDEs of two dimensions,
we cannot solve EOM except for a few cases
such as the Nambu-Goto strings in Minkowski spacetime,
where the EOM are reduced to wave equations in two dimensions
with constraint equations.

A way to make the EOM tractable is to assume symmetry.
The trajectory of an extended object, which we call {\wv},
is a submanifold embedded in the spacetime manifold $\M$.
Assuming symmetry on the geometry of the {\wv} ,
we can simplify the EOM.
In particular, in the case when the {\co} symmetry exists, 
the EOM are reduced to ODEs.
Examples are seen in stationary strings
\cite{P34_Frolov:1988zn,P34_deVega:1993rm,P34_Larsen:1994ah,P34_Larsen:1995bp,
  P34_deVega:1996mv,P34_Frolov:1996xw,P34_Ogawa:2008qn}
and branes \cite{P34_Kubiznak:2007ca}, 
and {\co} strings
\cite{P34_Ishihara:2005nu,P34_Koike:2008fs,Kozaki:2009jj,
 Igata:2009fd,Igata:2009dr}.

A {\co} {\wv}  $\Sigma$ of $m$ dimensions 
is foliated by $(m-1)$-dimensional
orbits of a group $G$ which consists of isometries of $\M$.
It is apparent that 
$\Sigma$ is homogeneous along the $(m-1)$-dimensional orbits.
For a {\co} string, 
its two-dimensional {\wv}  is foliated by 
one-dimensional orbits of $G$,
so that the group $G$ is one-dimensional, 
and hence there is no variety on the structure of $G$. 
For higher dimensional {\co} objects,
the structures of the groups $G$ which act on the 
homogeneous orbits have a richer variety. 
For example, in the case of two-dimensional groups,
Abelian and non-Abelian groups can act on the orbits.

For {\co} strings, 
the Nambu-Goto equation is reduced to the geodesic equations on the
orbit space, $\M/G$.
The metric $\tilde{h}$ which appears in the geodesic equations is
clearly identified as the one of the form $\tilde{h} = |\bxi| \, h$, 
where $h$ is the metric determined by the requirement that
the projection $\M \to \M/G$, which identifies the points on each
orbit of $G$, be a Riemannian submersion,
and $|\bxi|$ is the norm of the Killing vector $\bxi$ generating the 
group $G$ \cite{P34_Ishihara:2005nu}.
The clarification of the metric structure in relation to $G$ make it
possible to study the integrability of the geodesic equations.
The present authors found exact solutions for all of the {\co} strings
in Minkowski spacetime \cite{Kozaki:2009jj}.

For higher dimensional {\co} objects,
we may also expect the reduction of the Nambu-Goto equation to the
geodesic equations. 
In the case that $G$ is Abelian,
Kubiznak et al showed that the reduction of the equations of motion
occurs in the higher dimensional Kerr-NUT-(A)dS spacetime
\cite{P34_Kubiznak:2007ca}.
However, it is not clear that the same reduction occurs in general.
The structure of the metric which appear in the geodesic
equations is not deeply understood.

In this paper, we study {\co} membranes,
$(2+1)$-dimensional {\wv}s embedded in the spacetime,
and give a general formulation of reducing their Nambu-Goto equations
to a geodesic problem on the orbit space. 
We also give a thorough classification of the {\co} membranes in 
Minkowski spacetime. 
A careful treatment is necessary in the classification 
because different symmetry groups $G$, which are subgroups of the
isometry group of the spacetime $\M$,
could give essentially the same solution to the Nambu-Goto equation.
For example, in Minkowski spacetime,
a subgroup acting on the $x$-$y$ plane
and another acting on the $y$-$z$ plane should be identified
because the orbits are equivalent geometrically. 
This identification is achieved by an isometry, 
a rotation around $y$-axis,  which maps the one plane to the other. 
Using these identification by isometries,
we can classify isometry subgroups in a given spacetime.
After the classification of the subgroups in Minkowski spacetime,  
we choose one subgroup for a {\co} membrane, 
as an example, and give solutions to the EOM.

In the next section,
we show that the Nambu-Goto equations for the {\co}
membranes are reduced to the geodesic equations in the orbit space. 
The structure of the metric used in the geodesic equations are also 
clarified. 
In Sec.~\ref{sec:classification}, we discuss the classification of
{\co} symmetry for membranes.
As an example, we carry out the classification in Minkowski spacetime
in Sec.~\ref{sec:minkowski}.
After the classification, we take a particular {\co} symmetry
and solve the Nambu-Goto equations in Sec.~\ref{sec:solution}.
Finally, we summarize and discuss the results in
Sec.~\ref{sec:summary}.

\section{Reduction of equations of motion of {\co} membranes}
\label{sec:c1membrane}
We shall give a general formulation for
reducing the Nambu-Goto equations of {\co} membranes.
We first give a setup for {\co} membranes. 
There exit two cases, where the action of the symmetry group $G$ on
the orbits is simply or multiply transitive. 
Then we present the method of reducing the equations of motion 
in each case.

A membrane has a trajectory which is a three-dimensional surface 
embedded in a spacetime manifold $\M$. 
Let $\Isom\M$ be the isometry group of $\M$. 
A membrane is {\co} if its {\wv}  $\Sigma$ 
is foliated by two-dimensional orbits of a subgroup $G$ of $\Isom\M$.
We assume that the orbits are non-null. 
Let $\pi$ be the projection $\M\to\M/G$ 
which identifies the points on each orbits of $G$ in $\M$.
By the projection $\pi$, the spacetime manifold $\M$ is reduced to the 
orbit space $\M/G$, and the {\wv}  $\Sigma$ of {\co}
membrane is reduced to a curve $\C{}$ in $\M/G$. 
Thus the {\wv}  $\Sigma$ is given as a preimage
$\pi^{-1}(\C{})$
and is completely determined by the curve $\C{}$.
In the following subsections, we will show that the curve $\C{}$ is a
geodesic on $\M/G$, 
endowed with an appropriate metric, 
when the membrane is governed by the Nambu-Goto action.

Before proceeding, let us discuss the dimensionality of $G$.
The action of $G$ on the orbits may be 
simply transitive or multiply transitive.
In the simply transitive case, 
the isotropy subgroups are trivial 
and the dimensionality of $G$ is equal to that of the orbits, 
$\dim G = 2$. 
In the multiply transitive case,
$G$ includes a non-trivial isotropy subgroup, so that $\dim G > 2$.
On the other hand, the maximal dimensionality of the isometry group
acting on a two-dimensional surface is three, 
then we have $\dim G = 3$, and the each orbit is a 
space of constant curvature.

\subsection{The case $\boldsymbol{\dim G = 2}$}
\label{subsec:dimG_2}
Let $\pair{\bxi_1}{\bxi_2}$ be a pair of Killing vectors 
which are generators of $G \subset \Isom\M$. 
The Killing vectors $\bxi_I\ (I = 1, 2)$ are tangent to the orbits
and constitute a basis of the Lie algebra $\g$ of $G$.
It is known that there are only two distinct two-dimensional Lie
algebras, commutative and non-commutative.
With an appropriate choice of the basis $\pair{\bxi_1}{\bxi_2}$,
the Lie bracket is given by 
\begin{equation}
  \commutator{\bxi_1}{\bxi_2} =
  \begin{cases}
    0 & \text{($\g$ is commutative)} \\
    \bxi_1 & \text{($\g$ is non-commutative)}.
  \end{cases}
  \label{eq:commu2}
\end{equation}
For commutative $\g$'s, 
it was shown that the Nambu-Goto equations in a particular spacetime
are reduced to the geodesic equations \cite{P34_Kubiznak:2007ca}.
It has not been known whether such a reduction is possible
for non-commutative $\g$'s. 
In the following, we show that this is also true.

\subsubsection{Coordinate system in $\M$}
\begin{figure}[ht]
  \centering
  \includegraphics{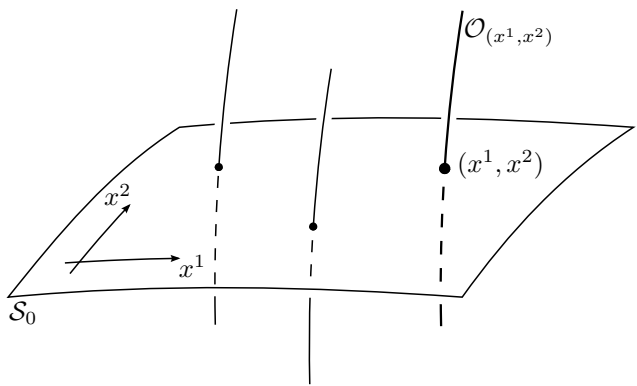}
  \caption{The surface $\S{0}$ and the orbits of $G$.
    The orbits are depicted as curves though they are 
    actually two-dimensional.
    Each orbit has only one intersection with $\S{0}$.}
  \label{fig:S_0}
\end{figure}

We shall provide $\M$ with a coordinate system by making use of the
group action of $G$ on $\M$. 
First, we consider a two-dimensional surface $\S{0}$ 
such that each orbit of $G$ intersects with $\S{0}$ once. 
Introducing a coordinate system $(x^1, x^2)$ on $\S{0}$, 
we can specify the orbit by the point $(x^1,x^2)$
of intersection with $\S{0}$, 
which we will denote by $\Orb{(x^1, x^2)}$ (see Fig.~\ref{fig:S_0}).

Next, we consider the action of an element $g$ of $G$ on
the points of $\S{0}$. 
Since $G$ does not admit fixed points,
each point on $\S{0}$ is necessarily moved along its orbit
except in the case that $g$ is the identity element $e$ of $G$.
The moved points form a new surface which does not intersect with
$\S{0}$. 
We denote this surface by $\S{g}$
and consider a family of the surfaces $\{\S{g}\}:=\{\S{g}|g \in G\}$
where $\S{e} = \S{0}$.
It is clear that the surfaces of $\{\S{g}\}$ fill the spacetime
without intersecting with each other.

\begin{figure}[ht]
  \centering
  \includegraphics{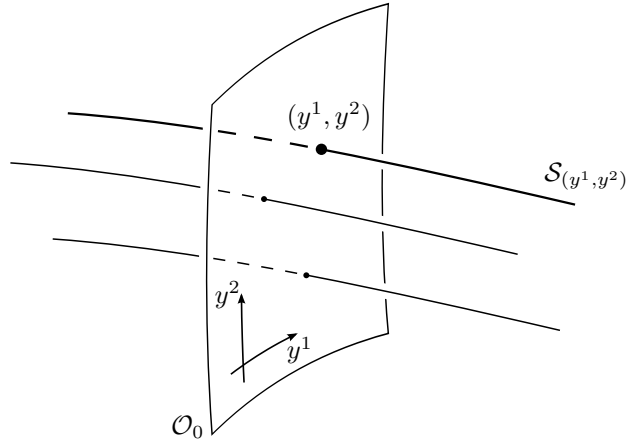}
  \caption{The orbit $\Orb{0}$ and the surfaces of $\{\S{g}\}$.
    Two-dimensional surfaces are depicted as curves.
  }
  \label{fig:O_0}
\end{figure}

Let us now choose an orbit, which we denote by $\Orb{0}$.
All the surfaces of $\{\S{g}\}$ cross the orbit $\Orb{0}$ at
different points,
and hence the surfaces are specified by the intersections on $\Orb{0}$.
Let $(y^1,y^2)$ be an internal coordinate system of $\Orb{0}$.
We can denote by $\S{(y^1, y^2)}$ the surface which intersects with
$\Orb{0}$ at $(y^1, y^2)$ (see Fig.~\ref{fig:O_0}).

Now that we have two different ways of filling $\M$:
one is with the orbits of $\{\Orb{(x^1, x^2)}\}$ and the other is
with the surfaces of $\{\S{(y^1, y^2)}\}$,
we can specify a point of $\M$ by the orbit $\Orb{(x^1,x^2)}$
and the surface $\S{(y^1,y^2)}$ on which the point lies.
Using the parameters of the orbit and the surface,
we can assign the coordinates $(x^1, x^2, y^1, y^2)$ to the point.
This coordinate system is convenient for studying the EOM of {\co}
membranes.

\subsubsection{Metric}
To describe the metric, 
let us introduce an invariant dual basis on each orbit, 
which is possible when the group action of $G$ on
the orbits is simply transitive. 
Let $\{\chi^1, \chi^2\}$ be an invariant dual basis on $\Orb{0}$,
which satisfy
\begin{equation}
 \mathcal{L}_{\bxi_I} \chi^J = 0,~~(I,J=1,2) \label{eq:invariant}
\end{equation}
where $\mathcal{L}_{\bxi}$ represents the Lie derivative along a
vector field $\bxi$.
With respect to the coordinate system $(y^1,y^2)$ on $\Orb{0}$,
$\chi^I$ is written as
\begin{equation}
 \chi^I = \chi^I{}_i(y^1,y^2) \,dy^i.
\end{equation}
Considering $y^1$ and $y^2$ as the spacetime coordinates, 
we can extend $\chi^I$ to 1-forms in $\M$ satisfying
Eq.~\eqref{eq:invariant}.

Using the invariant dual basis $\{\chi^1,\chi^2\}$, 
we can write the spacetime metric as
\begin{equation}
 ds^2 = g_{pq} dx^p dx^q + 2 g_{pI} dx^p \chi^I + g_{I\!J} \chi^I
 \chi^J. 
\end{equation}
Here, $g_{pq}, g_{pI}$ and $g_{I\!J}$ are functions of $x^1$ and $x^2$
only, 
which is due to the Killing equations
\begin{equation}
  \mathcal{L}_{\bxi_I} g = 0, 
\end{equation}
and Eq.~\eqref{eq:invariant}.
For later convenience, we write the metric as follows:
\begin{equation}
 ds^2 =   h_{pq} dx^p dx^q
        + g_{I\!J}(\chi^I + N^I{}_p dx^p )(\chi^J + N^J{}_q dx^q),
         \label{eq:metric}
\end{equation}
where 
\begin{align}
 &g_{I\!J} N^J{}_p = g_{Ip}, \\
 &h_{pq} = g_{pq} - g_{I\!J}N^I{}_{p}N^J{}_{q}.
\label{projection}
\end{align}

\subsubsection{Equations of motion}
\begin{figure}[ht]
  \centering
  \includegraphics{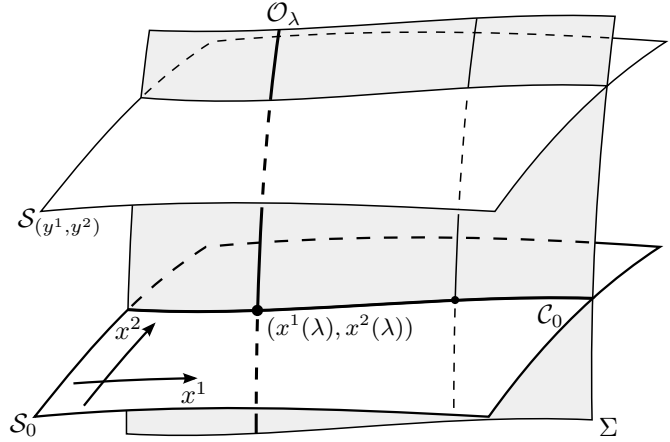}
  \caption{%
    The curve $\C{0}$ on the surface $\S{0}$ and the orbits
    $\Orb{\lambda}$ determine the {\wv}  $\Sigma$. The dimensions
    of $\Sigma$ is three, and of $\Orb{\lambda}$ is two. }
  \label{fig:C_0}
\end{figure}

When we identify the orbits with the points on $\S{0}$,
the {\wv}  $\Sigma$ is reduced to a curve on $\S{0}$.
We denote this curve by $\C{0}(\lambda)$:
\begin{equation}
  \C{0}: \R \ni \lambda \mapsto (x^1(\lambda),x^2(\lambda)) \in \S{0}.
\end{equation}
Then we can label the foliating orbits with the parameter $\lambda$:
\begin{equation}
 \Orb{\lambda} := \Orb{(x^1(\lambda), x^2(\lambda))}.
 ~~(\text{see Fig.~\ref{fig:C_0}})
\end{equation}

A point on $\Sigma$ is specified by the orbit $\Orb{\lambda}$ and
the surface $\S{(y^1, y^2)}$ on which the point lies.
Then the set of parameters $(\lambda, \,y^1, y^2)$
is considered as a coordinate system on $\Sigma$.
With this coordinate system,
the embedding of $\Sigma$ into $\M$ is given by
\begin{equation}
  (\lambda, \, y^1, y^2) \mapsto (x^1(\lambda), x^2(\lambda), y^1, y^2),
\end{equation}
and 
the Nambu-Goto action is given by the three-volume integral
\begin{equation}
  S = \int_{\Sigma} \!\!\sqrt{|\gamma|}\,\, d\lambda \,dy^1dy^2
    = \int_{\C{0}}
        \!\left\{
          \int_{\Orb{\lambda}} \!\!\sqrt{|\gamma|}
          \,\,dy^1dy^2
        \right\}
      d\lambda
      \label{eq:Nambu-Goto-Original}
\end{equation}
where $\gamma$ is the determinant of the induced metric on $\Sigma$,
and hence $\sqrt{|\gamma|}$ is the volume of the parallelepiped
spanned by the coordinate basis
$(\partial_{\lambda}, \partial_1, \partial_2 )
:= (\partial/\partial \lambda, \partial/\partial y^1, \partial/\partial y^2)$.

\begin{figure}[ht]
  \centering
  \includegraphics{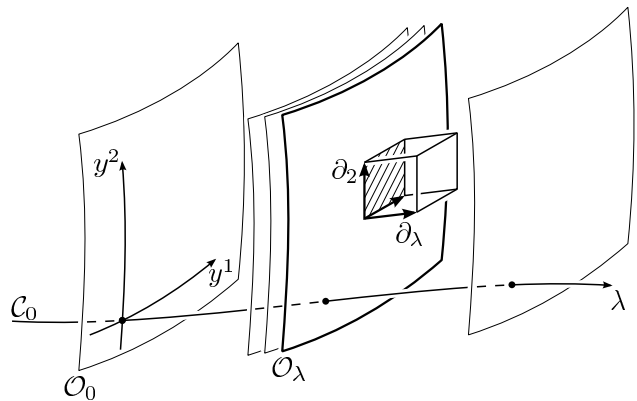}
  \caption{The parallelepiped spanned by the 
    coordinate basis: $(\partial_{\lambda}, \partial_1, \partial_2)$.
    The shaded parallelogram is spanned by $\partial_1$ and
    $\partial_2$.
    We should note that $\partial_{\lambda}$ is not necessarily
    perpendicular to the parallelogram.
  }
  \label{fig:parallelepiped}
\end{figure}

The volume of the parallelepiped is given by
a product of the area of the base and the height from the base. 
Considering the parallelogram spanned by $\partial_1$ and $\partial_2$
as the base (see Fig.~\ref{fig:parallelepiped}),
we obtain the area of the base as 
\begin{equation}
  \bigl|\det\chi^I{}_j\bigr| \sqrt{\bigl|\det g_{I\!J}\bigr|},
\end{equation}
and the height of the parallelepiped, which is given by the magnitude of the normal 
component of $\partial_{\lambda} = \dot{x}^p\partial/\partial x^p$ to the base
as
\begin{equation}
  \sqrt{\bigl|h_{pq} \,\dot{x}^p \,\dot{x}^q \bigr|},
  \label{eq:height}
\end{equation}
where the dot denotes the derivative with respect to $\lambda$.
Thereby the volume is written as
\begin{equation}
\sqrt{|\gamma|}
 = \bigl|\det\chi^I{}_j\bigr| \sqrt{\bigl|\det g_{I\!J}\bigr|}
   \sqrt{\bigl|h_{pq} \,\dot{x}^p \,\dot{x}^q \bigr|}
 = \bigl|\det\chi^I{}_j\bigr|
   \sqrt{\bigl| (\det g_{I\!J}) \, h_{pq} \, \dot{x}^p \, \dot{x}^q \bigr|}.
\end{equation}
Noting that $\chi^I{}_j$ are functions of $y^1$ and $y^2$
and that $g_{I\!J}$ and $h_{pq}$ are functions of $\lambda$,
we can write the Nambu-Goto action \eqref{eq:Nambu-Goto-Original}
as 
\begin{align}
  S
  &= \int_{\C{0}}
       \left\{
         \sqrt{\bigl|
           (\det g_{I\!J})\, h_{pq} \,\dot{x}^p \,\dot{x}^q
         \bigr|}
       \int_{\Orb{\lambda}}
             \!\!\bigl| \det\chi^I{}_j \bigr| 
             \,dy^1 dy^2
       \right\}
     d\lambda \\
  &= \int_{\Orb{}} \,\bigl|\det\chi^I{}_j\bigr| \,dy^1 dy^2
     \int_{\C{0}}
       \!\!\sqrt{\bigl|
                   (\det g_{I\!J}) \, h_{pq} \, \dot{x}^p \, \dot{x}^q 
                 \bigr|}
     \,d\lambda. \label{eq:separation}
\end{align}
Here, we have used the fact that the integration over the orbit 
$\Orb{\lambda}$ does not depend on $\lambda$. 
Integrating out the variables $y^1$ and $y^2$,
we can reduce the Nambu-Goto action as follows:
\begin{equation}
  S \propto \int_{\C{0}}
        \!\!\sqrt{\bigl|
                    (\det g_{I\!J})\, h_{pq} \,\dot{x}^p \,\dot{x}^q
                  \bigr|}
      \,d\lambda. \label{eq:reducedaction}
\end{equation}
This is identical to the action of a particle moving on the surface
$\S{0}$ with the metric $(\det g_{I\!J}) h_{pq}$.
The original problem has therefore been reduced to that of 
finding a geodesic on $\S{0}$.

We have derived the equations of motion on the surface $\S{0}$.
It is also possible to do so on 
other surfaces of $\{\S{g}\}$. 
The surfaces can be mapped to each other by the action of 
$g$ (or $g^{-1}$)
and lead to the same form of \eqref{eq:reducedaction}.
The reductions on the surfaces of $\{\S{g}\}$ are put together 
by identifying the surfaces with the orbit space $\M/G$ by 
\begin{equation}
  \S{g} \ni (x^1, x^2) \leftrightarrow \Orb{(x^1, x^2)} \in \M/G.
\end{equation}
We can then conclude that the equations of motions are reduced to
the geodesic equations on $\M/G$.
If we consider $(x^1, x^2)$ be a coordinate system of $\M/G$,
the projection map $\pi: \M \to \M/G$ is explicitly given as
\begin{equation}
  \pi : \M \ni (x^1, x^2, y^1, y^2) \mapsto (x^1, x^2) \in \M/G.
  \label{eq:projection}
\end{equation}
Let $(x^1(\lambda), x^2(\lambda))$ be a geodesic in $\M/G$.
The solution of the membrane is given as a preimage:
$\pi^{-1}(x^1(\lambda), x^2(\lambda))
= (x^1(\lambda), x^2(\lambda), \,y^1, \,y^2)$.
By the use of the projection $\pi$,
we can naturally induce a metric on $\M/G$ in order that $\pi$ is a 
Riemannian submersion.
Such an induced metric is given as the symmetric tensor $h_{pq}$ of
\eqref{projection}.
Weighting $h_{pq}$ with $\det g_{IJ}$,
we obtain the metric used in the geodesic action
\eqref{eq:reducedaction} for the {\co} membrane.

\subsection{The case $\boldsymbol{\dim G = 3}$}
\label{subsec:dimG_3}
In the case $\dim G = 3$,
the orbits are two-dimensional spaces of constant curvature.
The metric of the orbit with constant curvature $K$ 
can be written in the form:
\begin{align}
  d\sigma^2
  =
  R^2\Bigl\{(dy^1)^2 + \epsilon\, F^2(y^1)\,(dy^2)^2\Bigr\}
\end{align}
with
\begin{equation}
 \epsilon :=
 \begin{cases}
   +1 &\text{(spacelike orbits)} \\
   -1 &\text{(timelike orbits)}
 \end{cases}
\end{equation}
and
\begin{equation}
  F(y^1) := 
  \begin{cases}
   \sin y^1 & (K > 0)\\
   y^1 & (K = 0)\\   
   \sinh y^1 & (K < 0)
  \end{cases},
\end{equation}
where $R$ is a constant.
By using coordinates $x^1$ and $x^2$ 
which are constant on each orbit, 
we can write the spacetime metric as \cite{P34_Goenner:1970,P34_ExactSolution}:
\begin{align}
  ds^2 = e^{2 \lambda} (dx^1)^2 - \epsilon e^{2 \nu} (dx^2)^2
        + R^2 \Bigl\{(dy^1)^2 + \epsilon F^2(y^1) (dy^2)^2\Bigr\},
        \label{eq:metric_3dim}
\end{align}
where $\lambda, \nu$ and $R$ are functions of $x^1$ and $x^2$.
We can consider $(x^1, x^2)$ as a coordinate system of the orbit space 
$\M/G$ and the projection $\pi: \M \to \M/G$ is again given by
Eq.~\eqref{eq:projection}. 
We see that 
\begin{align}
  ds_{\M/G}^2 := e^{2 \lambda} (dx^1)^2 - \epsilon e^{2 \nu} (dx^2)^2
        \label{eq:metric_orbit}
\end{align}
is the metric on $\M/G$ with $\pi$ being 
a Riemannian submersion.
Following the same derivation presented in the case $\dim G = 2$,
the Nambu-Goto action is separated as 
\begin{equation}
 S = \int_{\Orb{}} \bigl|\Sigma\bigr| \,dy^1 dy^2
     \,
     \int_{\C{}} \sqrt{%
       \left|R^4
         \Bigl\{%
           e^{2 \lambda} (\dot{x}^1)^2
           - \epsilon e^{2 \nu} (\dot{x}^2)^2
         \Bigr\}
       \right|}\,d\lambda.
\end{equation}
Integrating out the variables $y^1$ and $y^2$, 
we obtain a reduced action
\begin{equation}
  S \propto \int_{\C{}} \sqrt{%
       \left|R^4
         \Bigl\{%
           e^{2 \lambda} (\dot{x}^1)^2
           - \epsilon e^{2 \nu} (\dot{x}^2)^2
         \Bigr\}
       \right|}\,d\lambda.
\end{equation}

Therefore the problem is reduced to solving the geodesic equations
on $\M/G$ with the weighted metric
\begin{equation}
 \widetilde{ds}_{\M/G}^2
  := R^4 \,ds_{\M/G}^2
  = R^4 \Bigl\{%
           e^{2 \lambda} (dx^1)^2 - \epsilon e^{2 \nu} (dx^2)^2
        \Bigr\}.
        \label{eq-orbitmetric}
\end{equation}
We finally note the Lie algebra $\g$ of $G$.
In the case $\dim G = 3$, the action of $G$ is described by three
Killing vector fields, $\bxi_I~(I = 1,2,3)$.
The triple $\triple{\bxi_1}{\bxi_2}{\bxi_3}$ represents a basis of $\g$.
The commutation relations are those of the two-dimensional
spaces of constant curvature, 
which are listed in Table \ref{table:3dim_Lie_alg} in terms of
the Bianchi classification of the three-dimensional Lie algebras.
As seen in Table \ref{table:3dim_Lie_alg},
the Lie algebras of Bianchi \VIzero{} and \VIIzero{} have
two-dimensional commutative subalgebras spanned by $\bxi_1$ and
$\bxi_2$. 
As for the Bianchi {\VIII}, taking new bases
\begin{equation}
  \bxi'_1 := \bxi_2 - \bxi_3,~~\bxi'_2 := \bxi_1,
\end{equation}
which satisfy
\begin{equation}
  \commutator{\bxi'_1}{\bxi'_2} = \bxi'_1,
\end{equation}
we find that the Bianchi {\VIII} also have a two-dimensional 
solvable subalgebra. 
Since the Lie algebras of Bianchi {\VIzero}, {\VIIzero} and  {\VIII}
include two-dimensional subalgebras,
the groups associated with these Lie algebras include two-dimensional
subgroups, 
whose actions on the orbits are simply transitive.
Then, for the groups of Bianchi {\VIzero}, {\VIIzero} and {\VIII},
the reduction of the EOM can be explained in the case $\dim G = 2$.

\begin{table}
  \setlength{\tabcolsep}{1em}
\begin{tabular}{lrrrr}
 \hline\hline
 spaces of constant curvature & Bianchi type
 &$\commutator{\bxi_1}{\bxi_2}$& $\commutator{\bxi_2}{\bxi_3}$ & $\commutator{\bxi_3}{\bxi_1}$ \\
 \hline
 Euclid space $E^{2}$   & \VIIzero & $0$      & $- \bxi_1$ & $-\bxi_2$ \\
 sphere $S^2$            & \IX    & $-\bxi_3$ & $-\bxi_1$ & $-\bxi_2$ \\
 hyperbolic space $H^2$  & \VIII & $\bxi_3$  & $-\bxi_1$ & $-\bxi_2$ \\
 Minkowski spacetime $E^{1,1}$   & \VIzero & $0$ & $\bxi_1$ & $- \bxi_2$ \\
 de Sitter spacetime $\dS$ & \VIII  & $\bxi_3$ & $-\bxi_1$ & $-\bxi_2$ \\
 anti-de Sitter spacetime $\AdS$ & \VIII   & $\bxi_3$ & $-\bxi_1$ & $-\bxi_2$ \\
 \hline\hline
 \end{tabular}
 \caption{The commutation relations of three Killing vectors 
   $\triple{\bxi_1}{\bxi_2}{\bxi_3}$
   of the two-dimensional spaces of constant curvature.
   }
\label{table:3dim_Lie_alg}
\end{table}

\section{General classification method for {\co} membranes}
\label{sec:classification}
The {\co} symmetries can be assumed in the spacetimes whose isometry 
groups admit subgroups $G$ with two-dimensional orbits.
In a given such spacetime, different $G$ may give the  {\co} membranes
which are essentially the same. 
To discard such redundancy, 
we shall introduce the notion of geometrical equivalence of {\wv}s 
and present the method of classifying
{\co} membranes up to the equivalence. 

Let $\Sigma$ and $\Sigma'$ be {\wv}s. 
We say that they are geometrically equivalent 
if there is an isometry $\phi$ on ${\M}$ 
which maps $\Sigma$ onto $\Sigma'$.
Suppose further that one of such {\wv}s, $\Sigma$, 
is of {\co} with symmetry group $G$. 
Then $\Sigma'$ is of {\co} with symmetry group 
$\phi G \phi^{-1}$. 
To see this, let $\Orb{\lambda}$ be the orbits of $G$ which comprise a
foliation of $\Sigma$.  
Then for each pair of points $p$ and $p'$ on $\phi(\Orb{\lambda})$, 
there exists $g\in G$ such that $g(\phi^{-1}(p))=\phi^{-1}(p')$ 
which implies 
$\phi g\phi^{-1}(p)=p'$. 
Thus $\phi(\Orb{\lambda})$ comprise a homogeneous foliation of $\Sigma'$
with symmetry group $\phi G\phi\inv$.

It is natural to introduce an equivalence relation for subgroups $G$
and $G'$ of $\Isom\M$:  
\begin{equation}
  G \sim G' \Leftrightarrow
  {}^\exists \phi \in \mathrm{Isom}\,\M
  ~~\text{s.t.}~ G' = \phi G \phi^{-1}.
  \label{eq:equiv} 
\end{equation}
Then subgroups $G$ and $G'$ define geometrically equivalent {\wv}s 
if and only if they are equivalent. 
Our task is to find out the equivalence classes of the relation 
\eqref{eq:equiv}, 
or the conjugacy class of the subgroups $G$ of  $\Isom \M$.

In the actual classification procedure,
it is more convenient to work with the Lie algebra $\g$ of $G$, 
which consists of Killing vector fields on $\M$.
Because the conjugation for $g\in G$ by $\phi\in\Isom\M$, 
$g\mapsto \phi g\phi\inv$, 
induces the pushforward $\bxi\mapsto\pf{\phi}{\bxi}$ for $\bxi\in\g$. 
The equivalence relation on the symmetry group $G$ 
induces that on the symmetry Lie algebra $\g$: 
\begin{align}
  \g\sim\g' \Leftrightarrow 
  {}^\exists \phi \in \mathrm{Isom}\,\M
  ~~\text{s.t.}~ \g' = \phi_*\g. 
  \label{eq:equiv-alg} 
\end{align}
Thus we shall classify the Lie subalgebras $\g$ of the Lie algebra of
$\Isom\M$, up to the equivalence relation \eqref{eq:equiv-alg}.

We further would like to derive a basis for each classified symmetry
Lie algebra $\g$, so that it is convenient in applications. 
Let $V:=(\bxi_1,...,\bxi_{\dim G})$ a basis  for a symmetry algebra $\g$. 
The bases $V$ and $V' := (\bxi'_1,...,\bxi'_{\dim G})$ give the same
$\g$ when each element of $V'$ is a linear combination of the elements
of $V$: $\bxi'_i = {A_i}^j \bxi_j$, $A_i{}^j\in \GL{\dim G}{\R}$.
Then the classification of all {\co} membrane in a given 
spacetime $\M$ reduces to that of the bases for the symmetry algebras
under the equivalence relation 
\begin{align}
  V \sim  V' \Leftrightarrow
  {}^\exists \phi \in \mathrm{Isom}\,\M, 
  {}^\exists {A_i}^j \in \GL{\dim G}{\R}
  ~~\text{s.t.}~ \bxi'_i = {A_i}^j \pf{\phi}\bxi_j.
  \label{eq:equiv_gendim} 
\end{align}
A concrete procedure to get a set of class representatives is the
following.

Step 1. Choose an abstract Lie algebra $\g$ of the symmetry group $G$. 
It must be one of the following six: 
\begin{align}
&\text{$\R^2$, two-dimensional non-commutative algebra}, 
\nonumber \\
&\text{Bianchi types {\VIzero}, {\VIIzero}, {\VIII} and {\IX}}. 
\label{eq-listliealg}
\end{align}
As discussed in Sec.~\ref{sec:c1membrane}, 
the above are the only Lie algebras that allow two-dimensional
orbits. 
Furthermore, as mentioned at the end of Sec.~\ref{sec:c1membrane}, 
one can eliminate 
Bianchi types {\VIzero}, {\VIIzero} and {\VIII} from the list
\eqref{eq-listliealg},  
because they are the special cases of $\R^2$ and two-dimensional
non-commutative algebra.
However, here we retain them so as to include all possible cases 
that the orbits are spaces of constant curvature and the metric on the
orbit space has the simple form \eqref{eq-orbitmetric}.

Step 2. Find a general set of Killing vector fields on $\M$, 
$V=(\bxi_1,...,\bxi_{\dim G})$, 
that satisfy one of the commutation relations \eqref{eq:commu2} and
those in Table \ref{table:3dim_Lie_alg} depending on the Lie algebra
chosen in Step 1.
Check that the orbit of $V$ is two-dimensional.

Step 3-$k$ ($k=1,...,\dim G$). 
Canonicalize $\bxi_k$. 
Namely, reduce $\bxi_k$ to a certain simple form by using the degrees
of freedom of the equivalence relation \eqref{eq:equiv_gendim} that
preserves $\bxi_l$ for $l<k$. 
We shall say that such $(\bxi_1,...,\bxi_k)$ has the canonical form. 
[We might sometimes rearrange the canonical form $(\bxi_1,...,\bxi_k)$
by  using $\GL{k}{\R}$ in order to make it look simpler as a whole.]
Finally, with $k=\dim G$, we obtain the canonical form of $V$.

\section{Classification in Minkowski spacetime}
\label{sec:minkowski}
We have obtained the general 
scheme to classify {\co} membranes in a given spacetime. 
In this section, we carry out the complete classification 
in Minkowski spacetime, which admits ten linearly independent Killing
vectors, 
\begin{equation}
  \begin{array}{rll}
    \bP{\mu} & ~~\kern-0.3em(\mu = t, \,x,\, y,\, z),
             & ~~\text{Translations}; \\
    \bK{i}   & ~~\kern-0.3em(i = x,\,y,\,z),
             & ~~\text{Lorentz boosts};\\
    \bL{i}   & ~~\kern-0.3em(i = x,\,y,\,z),
             & ~~\text{Rotations}.
  \end{array}  
\label{eq:standard}
\end{equation}
Any Killing vector $\bxi$ is written as a linear combination
of them, 
\begin{equation}
  \bxi = \alpha_{\mu} \bP{\mu} + \beta_{i} \bK{i} + \gamma_{i} \bL{i},
  \label{eq:arbitrary}
\end{equation}
where $\alpha_{\mu}$, $\beta_{i}$ and $\gamma_{i}$ are constants.

In Minkowski spacetime, 
we have an advantage that greatly simplifies the classification 
scheme because all canonical forms of Killing vector fields are
derived~\cite{P34_Ishihara:2005nu}. 
For any $\g$, we can assume that $\bxi_1$ is 
one of the canonical forms listed in Table \ref{table:7types} up to
scalar multiplication. 
Thus Step 3-1 is essentially done in advance.

\newcommand{\canonicalform}[2]{$a #1 + b #2$}
\begin{table}
  \setlength{\tabcolsep}{1em}
  \begin{tabular}{ll}
    \hline\hline
    Type & Canonical form \\
    \hline
    I   & \canonicalform{\bP{t}}{\bL{z}} \\
    II  & \canonicalform{(\bP{t} + \bP{z})}{\bL{z}} \\
    III & \canonicalform{\bP{z}}{\bL{z}} \\
    IV  & \canonicalform{\bP{z}}{(\bK{y} + \bL{z})}\\
    V   & \canonicalform{\bP{z}}{\bK{y}} \\
    VI  & \canonicalform{\bP{x}}{(\bK{y} + \bL{z})} \\
    VII & \canonicalform{\bK{z}}{\bL{z}} \\
    \hline\hline
    & ($a$, $b$: constants)
  \end{tabular}
  \caption{%
    Canonical forms of the Killing vectors in Minkowski spacetime.
    Any Killing vectors are transformed by isometries
    to these seven types.
    Killing vectors in the same canonical form with
    different pairs of constants $(a,b)$ cannot be transformed to
    each other.}
  \label{table:7types}
\end{table}

\subsection{Classification of two-dimensional Abelian symmetry groups}
\label{subsec:two-dim_abelian}
Let us choose the two-dimensional commutative algebra $\R^2$ as the
symmetry algebra $\g$. 
We would like to derive the equivalence classes of the set of
commuting pairs of Killing vector fields, 
\begin{equation}
 \Vc
 := \left\{%
      \pair{\bxi_1}{\bxi_2}\,|\,
      \commutator{\bxi_1}{\bxi_2} = 0
    \right\}.
\end{equation}

As discussed above, Step 3-1 is already carried out and 
we can take $\bxi_1$ as one of the canonical forms of Table
\ref{table:7types} up to scalar multiplication. 
We then divide $\Vc$ essentially into seven parts
$\Vc^{J}~~(J = \mathrm{I},~\mathrm{II}, \dots, \mathrm{VII}$)
depending on the canonical form of $\bxi_1$.
For example, $\Vc^{\mathrm{I}}$ is a set of the
commuting pairs of Killing vectors $\pair{\bxi_1}{\bxi_2}$ with
$\bxi_1$ being in the canonical form of Type I,
$\bxi_1 = a \bP{t} + b \bL{z}$,
and in this case $\bxi_2$ is written as linear combinations of
the ten Killing vectors \eqref{eq:standard} which commute with
$a \bP{t} + b \bL{z}$.
Next, we reduce the number of the Killing vectors \eqref{eq:standard}
contained in $\bxi_1$ and $\bxi_2$ by using isometries and
$\GL{2}{\R}$ actions on the pair $\pair{\bxi_1}{\bxi_2}$, 
so that $\bxi_1$ and $\bxi_2$ contains 
the smallest possible number of parameters.
A detailed calculation for one case is given in Appendix.

As a result, we obtain simple representatives in each of $\Vc^J$.
However,  we must be aware that 
two different $\Vc^J$'s may lead to equivalent pairs
$(\bxi_1,\bxi_2)$. 
We eliminate this redundancy and obtain a complete set of
canonical forms. The result is shown in Table \ref{table:result-2dim-comm}.
Any element of $\Vc$ falls into one of the equivalence classes of
these canonical forms. 

\begin{table}  \setlength{\tabcolsep}{1em}
\begin{tabular}{ll}
 \hline\hline
 $\bxi_1$ & $\bxi_2$~~ ($a$, $b$: constants) \\
 \hline\hline
 $\bP{t}$       & $a \bP{z} + b \bL{z}$ \\
 $\bP{t} + \bP{z}$
 & $a \bP{t} + b \bL{z}$,\ \ 
   $a \bP{z} + b (\bK{x} - \bL{y})$,\ \ 
   $a \bP{y} + b (\bK{x} - \bL{y})$ \\
 $\bP{z}$
 & $a \bP{t} + b \bL{z}$,\ \ 
   $a \bP{x} + b \bK{y}$,\ \ 
   $a \bP{x} + b (\bK{y} + \bL{z})$ \\
 $\bL{z}$ & $\bK{z}$ \\
 $\bK{y} + \bL{z}$ & $a \bP{z} + b (\bK{z} - \bL{y})$\\
 \hline\hline
\end{tabular}
 \caption{
   The representatives of the equivalence classes of $\Vc$:
   a set of commuting pairs of Killing vectors
   $\pair{\bxi_1}{\bxi_2}$.
   Those pairs that are connected by a simple rescaling 
   of $\bxi_2$ are equivalent, though having different $(a,b)$. 
}
\label{table:result-2dim-comm}
\end{table}

\subsection{Classification of two-dimensional  non-Abelian symmetry groups}
Let us choose the two-dimensional non-commutative algebra as the symmetry
algebra $\g$. 
This is the classification of 
$\Vs := \left\{%
      \pair{\bxi_1}{\bxi_2}\,|\,
      \commutator{\bxi_1}{\bxi_2} = \bxi_1
    \right\}
$. 
As in the commutative case, we can take $\bxi_1$ to be the seven types in
Table \ref{table:7types} and we reduce the degree of freedom in 
$\Vs$ by using $\phi\in\Isom\M$ and $A_i{}^j\in \GL{2}{\R}$ which
preserves the commutation relation.
The resulting canonical forms are listed in Table
\ref{table:result-2dim-sol}. 

\begin{table}
  \setlength{\tabcolsep}{1em}
  \begin{tabular}{ll}
    \hline\hline
    $\bxi_1$ & $\phantom{-}\bxi_2$~~ ($a$: constant) \\
    \hline\hline
    $\bP{t} + \bP{z}$  & $\phantom{-}\bK{z} + a \bP{x}$, $\bK{z} + a \bL{z}$ \\
    $\bK{y} + \bL{z}$ & $-\bK{x} + a \bP{z}$ \\
    \hline\hline
  \end{tabular}
  \caption{
      The representatives of the equivalence classes of $\Vs$:
      a set of non-commuting pairs of Killing vectors
      $\pair{\bxi_1}{\bxi_2}$. 
  }
  \label{table:result-2dim-sol}
\end{table}

\subsection{Classification of three-dimensional symmetry groups}
Let us discuss the case that the symmetry algebra $\g$ is three-dimensional. 
As was discussed in the previous section, 
$\g$ must be one of the Bianchi types {\VIzero}, {\VIIzero}, {\VIII}
and {\IX}.
The classification is for the triples of Killing vector fields 
$(\bxi_1,\bxi_2,\bxi_3)$ which satisfy either of the commutation
relations listed in Table \ref{table:3dim_Lie_alg}. 
The classification procedure for each Bianchi type is described in the
subsequent subsections. The result of the canonical forms for all Bianchi
types are shown in Table \ref{table:result-3dim}. 

\begin{table}
  \setlength{\tabcolsep}{1em}
  \begin{tabular}{llll}
    \hline\hline
    Bianchi type & $\bxi_1$ & $\bxi_2$& $\bxi_3$ \\
    \hline
    \VIzero & $\bP{t}$ & $\bP{z}$ & $\bK{z}$ \\
    \VIIzero& $\bP{z}$ & $\bP{x}$ & $\bL{y}$ \\
            & $\bK{y}+\bL{z}$ &$\bK{z}-\bL{y}$  & $\bL{x}$ \\
    \VIII   & $\bK{z}$ & $\bK{x}$ & $\bL{y}$ \\
    \IX     & $\bL{z}$ & $\bL{x}$ & $\bL{y}$ \\
    \hline\hline
  \end{tabular}
  \caption{
    The representatives of the equivalence classes 
    of triples of Killing vectors
    $\triple{\bxi_1}{\bxi_2}{\bxi_3}$ 
    that generate isometry group that has two-dimensional orbits.
  }
  \label{table:result-3dim}
\bigskip
\end{table}

\subsubsection{Bianchi {\VIzero}}
Bianchi type {\VIzero} algebra has a two-dimensional commutative
subalgebra $\h$. 
The subalgebra $\h$ must be equivalent to one of the Lie algebras
defined by the pairs in Table \ref{table:result-2dim-comm}. 
Then we should look for the third Killing vector $\bxi_3$ which satisfies the
commutation relations
\begin{equation}
 \commutator{\bxi_2}{\bxi_3} = \bxi_1, 
 \ \commutator{\bxi_3}{\bxi_1} = -\bxi_2. 
 \label{eq:commu3}
\end{equation}
where we take a linear combinations $\bxi_i$ if necessary. 
Reducing the general expression of 
$V=\triple{\bxi_1}{\bxi_2}{\bxi_3}$ 
by using the equivalence relation \eqref{eq:equiv_gendim} 
leads to the canonical form 
\begin{equation}
  V
  =
  \triple{\bP{t}}{\bP{z}}{\bK{z} + \alpha_{\mu}\bP{\mu}
    +\gamma\bL{z}}, 
 \label{eq:triples}
\end{equation}
where $\alpha_{\mu}$ and $\gamma$ are arbitrary constants.

Let us require that the symmetry group $G$ have the two-dimensional 
orbits. The tangent space spanned by $V$ at each point must be
two-dimensional.
The concrete expression of the Killing vector fields,  
\begin{equation}
  \bxi_1 = \partial_t, \ 
  \bxi_2 = \partial_z, \ 
  \bxi_3 = z\partial_t +  t \partial_z
           + \alpha_{x} \partial_{x}
           + \alpha_{y} \partial_{y}
           + \gamma (x \partial_y - y \partial_x), 
\end{equation}
in Cartesian coordinate basis implies that 
$\bxi_3$ cannot have the terms proportional to $\partial_x$ or 
$\partial_y$. 
Thus, we must have $\alpha_\mu=0$ and $\gamma=0$
so that 
\begin{equation}
  V
  = \triple{\bP{t}}{\bP{z}}{\bK{z}}.
  \label{eq:triple_VI0}
\end{equation}
The orbits of $\g$ are obviously
the planes parallel to the $t$-$z$ plane. 
We note that the Lie algebra spanned by the 
basis $V$ contains both of the commutative and non-commutative 
two-dimensional algebras, spanned by $\pair{\bP{t}}{\bP{z}}$ 
in Table \ref{table:result-2dim-comm} 
and $\pair{\bP{t} + \bP{z}}{\bK{z}}$
in Table \ref{table:result-2dim-sol}, respectively. 
Accordingly, the two-dimensional 
orbits of \eqref{eq:triple_VI0} can be considered as
those generated by $\pair{\bP{t}}{\bP{z}}$ or
by 
$\pair{\bP{t} + \bP{z}}{\bK{z}}$. 
The orbit is the two-dimensional Minkowski spacetime.

\subsubsection{Bianchi {\VIIzero}}
Let us consider the case of $\g$ being the Bianchi {\VIIzero}
algebra. 
Then $\bxi_1$ and $\bxi_2$ in 
$V=\triple{\bxi_1}{\bxi_2}{\bxi_3}$ 
commute.
Following the same procedures as in the case of Bianchi
{\VIzero}, we obtain two representatives, 
\begin{equation}
  V
  =
  \triple{\bP{z}}{\bP{x}}{\bL{y}}\ \text{and}\ 
  \triple{\bK{y} + \bL{z}}{\bK{z} - \bL{y}}{\bL{x}}. 
\end{equation}
In the first case,  
which we call type {\VIIzeroI}, 
the orbits are parallels to the $z$-$x$ plane and are intrinsically and
extrinsically flat.
In contrast, in the second case,
which we call type {\VIIzeroII}, 
the orbits are flat intrinsically but are embedded in $\M$ in an 
non-trivial way. 
Both types of embedding share common features:
intrinsic flatness and extrinsic homogeneity and isotropy. 
The type {\VIIzeroII} with 
non-trivial embedding of orbits seems worth further analysis. 
In Sec.~\ref{sec:solution}, 
we will clarify how the orbits of type {\VIIzeroII} 
are embedded in Minkowski spacetime, 
and will explicitly construct a solution of a {\co} Nambu-Goto 
membrane.

\subsubsection{Bianchi {\VIII}}
Let the symmetry algebra $\g$ be the Bianchi {\VIII} algebra. 
We start with 
$\bxi_1$ of the 
$V=\triple{\bxi_1}{\bxi_2}{\bxi_3}$ being one of the canonical form
in Table \ref{table:7types} (up to rescaling). 
Next, for the chosen $\bxi_1$,
we look for $\bxi_2$ which satisfies following relations 
\begin{align}
 \commutator{\bxi_1}{\commutator{\bxi_1}{\bxi_2}}
 &= \commutator{\bxi_1}{\bxi_3} =  \bxi_2, \\ 
 \commutator{\bxi_2}{\commutator{\bxi_1}{\bxi_2}}
 &= \commutator{\bxi_2}{\bxi_3} = - \bxi_1.
\end{align}
The third Killing vector $\bxi_3$ is obtained through the commutation
relation
\begin{equation}
 \bxi_3 = \commutator{\bxi_1}{\bxi_2}.
\end{equation}
We then have possible $V$. 
By using $\GL{3}{\R}$ that preserves the commutation relations,
we find that there is only one equivalence class 
represented by
\begin{equation}
  V
  = \triple{\bK{x}}{\bK{y}}{\bL{z}}.
 \label{eq:triple_VIII}
\end{equation}
The orbits are two-dimensional hyperboloids
or de Sitter spacetimes which are embedded in $E^{2,1}$ with
the equation
\begin{equation}
  - t^2 + x^2 + y^2 = \text{const.}
\end{equation}
As mentioned in subsection \ref{subsec:dimG_3},
the Lie algebra 
$\g$ 
spanned by $V$ includes a solvable subalgebra spanned by
$\pair{\bK{y} + \bL{z}}{-\bK{x}}$,
which is a special case in Table \ref{table:result-2dim-sol}.

\subsubsection{Bianchi {\IX}}
Let $\g$ be the Bianchi {\IX} algebra. 
As in the case of Bianchi {\VIII},
we first consider $\bxi_1$ to be in a canonical form in Table
\ref{table:7types}. 
We then look for $\bxi_2$ which satisfies
\begin{align}
 \commutator{\bxi_1}{\commutator{\bxi_1}{\bxi_2}}
 &= \commutator{\bxi_1}{-\bxi_3} =  -\bxi_2, \\ 
 \commutator{\bxi_2}{\commutator{\bxi_1}{\bxi_2}}
 &= \commutator{\bxi_2}{-\bxi_3} =  \bxi_1.
\end{align}
The third Killing vector $\bxi_3$ is obtained through the commutation
relation 
\begin{equation}
 \bxi_3 = \commutator{\bxi_1}{\bxi_2}.
\end{equation}
By using isometries and $\GL{3}{\R}$ actions, 
we then find that there is only one equivalence class 
represented by
\begin{equation}
  V
  =
  \triple{\bxi_1}{\bxi_2}{\bxi_3} 
  = \triple{\bL{z}}{\bL{x}}{\bL{y}}.
\end{equation}
The orbits are spheres centered at the origin.

\section{Exact solution for Type {\VIIzeroII} membrane}
\label{sec:solution}

Applying the results of subsection \ref{subsec:dimG_3},
we solve the Nambu-Goto equations for the {\co} membrane
whose {\wv}  has the symmetry of Bianchi type \VIIzero.
The symmetry algebra $\g$ has two possibilities: 
one is type {\VIIzeroI} generated by 
$\triple{\bP{z}}{\bP{x}}{\bL{y}}$ 
and the other is
type {\VIIzeroII} generated by 
$\triple{\bK{y} + \bL{z}}{\bK{z} - \bL{y}}{\bL{x}}$,

In the case of type \VIIzeroI,
the orbits are 
the $t=\text{const.}$ and $y=\text{const.}$ planes. 
The weighted metric of the orbit space, whose geodesics determine the
dynamics of the membrane, is flat. 
Then, {\co} membrane of type {\VIIzeroI} is a static plane or its
equivalents.

Hereafter, we concentrate on type \VIIzeroII. 
We follow the same conventions as in subsection \ref{subsec:dimG_3}:
the coordinates on the orbits are denoted by $(y^1, y^2)$, and
the orbits are distinguished by $(x^1, x^2)$.

\subsection{Embbeding of type \VIIzeroII ~orbits}
We begin with clarifying the embedding of the {\theorbits}:
$\vec{y}:=(y^1, y^2) \mapsto (t,x,y,z)$,
generated by the Killing vectors
\begin{equation}
  \triple{\bK{y}+\bL{z}}{\bK{z}-\bL{y}}{\bL{x}}.
  \label{eq:triple}
\end{equation}
Since the Killing vectors $\bK{y} + \bL{z}$ and $\bK{z} - \bL{y}$
commute with each other, 
we take them as a coordinate basis on the orbit:
\begin{equation}
  (\partial_{y^1},\partial_{y^2})
  = (\bK{y} + \bL{z}, \bK{z} - \bL{y}).
  \label{eq:coordinate_bases}
\end{equation}
In Cartesian coordinates 
$(t,x,y,z)$, 
Eqs.~\eqref{eq:coordinate_bases} are written as
\begin{align}
  \partial_{y^1}
  &= t_{,1}\partial_t + x_{,1} \partial_x + y_{,1} \partial_y + z_{,1} \partial_z
  = y(\partial_t - \partial_x) + (t + x) \partial_y, \\
  \partial_{y^2}
  &= t_{,2} \partial_t + x_{,2} \partial_x + y_{,2} \partial_y + z_{,2} \partial_z
  = z(\partial_t - \partial_x) + (t + x) \partial_z,
\end{align}
where $t$, $x$, $y$ and $z$ are considered as embedding functions,
namely functions of $y^i~(i = 1,2)$,
and the commas denote the differentiation with respect to $y^{i}$.
Comparing the coefficients of the coordinate basis,
we obtain equations of the embedding:
\begin{align}
  \dfrac{\partial t}{\partial y^1} = - \dfrac{\partial x}{\partial y^1}
  = y, &&
  \dfrac{\partial t}{\partial y^2} = - \dfrac{\partial x}{\partial y^2}
  = z, &&
  \dfrac{\partial y}{\partial y^1} =   \dfrac{\partial z}{\partial y^2}
  = (t + x), &&
  \dfrac{\partial z}{\partial y^1} =  \dfrac{\partial y}{\partial y^2}
  = 0.
\end{align}
These equations are readily solved as
\begin{align}
  t = \dfrac{a}{2}\, \vec{y}\!\cdot\!\vec{y} + \frac{a+b}{2}, ~~
  x = -t + a, ~~
  y = a y^1, ~~
  z = a y^2, \label{eq:orbit_embedding}
\end{align}
where $a$ and $b$ are arbitrary constants.
Eqs.~\eqref{eq:orbit_embedding} are equivalent to
the following implicit equations
\begin{align}
  - \left(t - \frac{b}{2}\right)^2 
  +\left (x + \frac{b}{2}\right )^2 + y^2 + z^2 =0, 
  \label{eq:lightcone_tx}\\
  t + x - a = 0. 
  \label{eq:nullplane_tx}
\end{align}
We see that each orbit is the cross section of 
a light cone \eqref{eq:lightcone_tx} and a null plane
\eqref{eq:nullplane_tx}.

With the null coordinates $u := t + x$ and $v := t - x$,
Eqs.~\eqref{eq:lightcone_tx} and \eqref{eq:nullplane_tx}
are written as
\begin{align}
  - u(v - b) + y^2 + z^2 = 0,  \label{eq:lightcone}\\
  u - a = 0. \label{eq:nullplane}
\end{align}
Therefore each orbit is a two-dimensional paraboloid 
\begin{equation}
  -a(v - b) + y^2 + z^2 = 0 \label{eq:paraboloid}
\end{equation}
on a null plane 
$u = a$.
Since the paraboloid is specified by the vertex, 
located at $(u, v, y, z) = (a, b, 0, 0)$,
we identify such each orbit with the point $(a, b)$ in the $u$-$v$ plane. 
Therefore the $u$-$v$ plane can be identified with the orbit space.
Hereinafter, we use the coordinate system $(u, v)$ of the orbit space
as the $(x^1, x^2)$ in subsection \ref{subsec:dimG_3}.

Combining the coordinate system on the orbit space $(u, v)$
and that on the orbit $(y^1, y^2)$, 
we make up a coordinate system $(u, v, y^1, y^2)$ in $\Minkowski$.
By the coordinate transformation between $(t,x,y,z)$ and $(u,v,y^1,y^2)$
given by \eqref{eq:orbit_embedding} with  $a = u$ and $b = v$, 
the metric of $\Minkowski$ is written as
\begin{equation}
  ds^2 = -du \, dv + u^2 d\vec{y}\,{}^2.
  \label{eq-VII0-2-metric}
\end{equation}
This form has the same structure of \eqref{eq:metric_3dim};
the first term is the metric on $\M/G$ such that 
the projection is a Riemannian submersion.

\subsection{Solutions for type \VIIzeroII ~membranes}

Following the results of subsection \ref{subsec:dimG_3},
the Nambu-Goto equations for the {\co} membrane is reduced to
the geodesic equations on $\M/G$ with the weighted metric 
\eqref{eq-orbitmetric} where $R^2=u^2$, 
\begin{equation}
  ds_{\M/G}^2 = u^4 (- du \, dv).
\end{equation}
In order to solve the geodesic equations,
we start with the action:
\begin{align}
  S = \int \left(\frac{\lagrangian}{N} - N\right) d\lambda, ~~~
  \lagrangian = - u^4 \dot{u} \,\dot{v},
\end{align}
where 
the dots denote the derivative with respect to the parameter $\lambda$,
and $N$ is a function of $\lambda$ which determines the parameterization
of the geodesic; indeed,
variation with $N$ leads
\begin{equation}
  - u^4 \dot{u} \, \dot{v} = -N^2.
  \label{eq:constraint}
\end{equation}
Variations with $u$ and $v$ give two conserved quantities
$C_u$ and $C_v$:
\begin{align}
  \frac{\dot{v}}{N} = \sqrt{5}\,C_{u}, \ \ 
  \frac{u^4 \dot{u}}{N} = C_{v}. 
  \label{eq-VII0-2-EOM}
\end{align}
The constraint condition \eqref{eq:constraint} gives
$
\sqrt{5}\, C_u C_v = 1. 
$ 
Then 
Eqs. \eqref{eq-VII0-2-EOM} are readily integrated as
\begin{equation}
  v(\lambda) = {C_u}^2 u^5(\lambda) + 2D,
\end{equation}
where $D$ is an arbitrary constant.
This curve on the $u$-$v$ plane describes the trajectory of the vertex
of the paraboloid \eqref{eq:paraboloid}.

\begin{figure}[ht]
  \centering
  \includegraphics{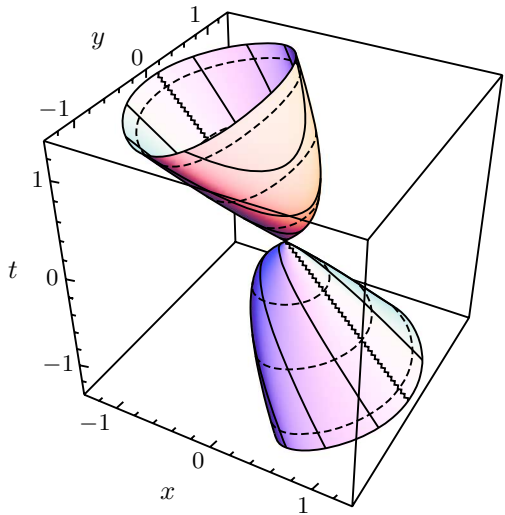}
  \caption{
    The {\wv}  of the membrane with $C_u = 1.1$.
    The $z$ direction is omitted.
    The lower {\wv}  shrinks to a point at $t = x = 0$,
    and the upper expands from there.
    The dashed lines represent $t = \text{const.}$ slices.
    The solid lines represent $t + x = \text{const.}$ slices,
    which also represent the foliating orbits of type \VIIzeroII.
    The jagged line is a singular orbit of dimension one with $u = 0$,
    which corresponds to the `cosmological singularity' of the
    intrinsic geometry.
  }
  \label{fig:solution}
\end{figure}

The embedding of the {\wv}  is implicitly written as
\begin{equation}
  -u(v - {C_u}^2 u^5 - 2D) + y^2 + z^2 = 0,
\end{equation}
or, equivalently,
\begin{equation}
  -(t - D)^2 + (x + D)^2 + y^2 + z^2 + C_u^2 (t + x)^6 = 0.
\end{equation}
Though the solution has two free parameters $C_u$ and $D$,
we can set $D=0$, i.e.,
\begin{equation}
  -t^2 + x^2 + y^2 + z^2 + C_u^2 (t + x)^6 = 0,
  \label{eq:solution}
\end{equation}
because the {\wv}  with $D \neq 0$ is identified 
with the one with $D = 0$ by using a translation for the null
direction.
As depicted in Fig.~\ref{fig:solution}, 
the $t = \text{const.}$ slices of the {\wv}  are closed;
then the solution represents a closed membrane, which
shrinks or expands.
In contrast, the slices with the null planes \eqref{eq:nullplane}
give the paraboloids of revolution 
\eqref{eq:paraboloid} 
 and hence are not closed.
Since the paraboloids are the orbits of the Killing vectors
\eqref{eq:triple},
the membrane is homogeneous and isotropic, actually flat,
on these null slices.

From 
Eqs. \eqref{eq-VII0-2-metric}, 
\eqref{eq:constraint} and 
\eqref{eq-VII0-2-EOM},  
the metric induced on the {\wv}  is 
written as
\begin{equation}
  ds_{\Sigma}^2 = - d\lambda^2 + u^{2} d \vec{y}\,{}^2,  ~~~~
  u(\lambda) \propto \lambda^{1/3},
  \label{eq:induced_metric}
\end{equation}
where we have chosen 
the parameterization of $\lambda$ so that 
\begin{equation}
  N(\lambda) = u^2(\lambda).
\end{equation}
The geometry on the {\wv}  is analogous to the flat
Friedmann-Lema\^itre-Robertson-Walker universe.
At the null line $u = 0$ on the {\wv} ,
i.e., $t + x = 0$, $y = z = 0$,
the scalar curvature of the induced metric \eqref{eq:induced_metric}
diverges.
Thus, the `cosmological singularity' is described by the null line
$u = 0$.
The orbits \eqref{eq:lightcone} and \eqref{eq:nullplane} generated by
the Killing vectors \eqref{eq:triple} are two-dimensional spacelike
surface, but is null line at $u = 0$.
Therefore the `cosmological singularity' is described by the singular
orbit.
We remark that the orbit is singular at $u=0$ but the {\wv} 
\eqref{eq:solution} itself is
smooth everywhere except at the origin of Minkowski spacetime. 
The embedding of the membrane is very similar to that of the brane
universe in five-dimensional anti-de Sitter space
\cite{Ishihara:2001qe}.

\section{Summary and discussion}
\label{sec:summary}
We have investigated the dynamics of {\co} membranes.
The three-dimensional {\wv}  of the {\co} membrane is 
foliated by two-dimensional orbits of the symmetry group $G$ 
that is a subgroup of the isometry group, $\Isom \M$, of the spacetime
$\M$.  
The symmetry suggests that the equations of motion are 
reduced to ordinary differential equations.
We have explicitly shown that the Nambu-Goto equations are
reduced to the geodesic equations in the orbit space, 
or the quotient space $\M/G$,
with a properly defined metric thereon.

In a highly symmetric spacetime, there exist a variety of symmetry 
groups $G$ that allow two-dimensional orbits.
We have proposed a classification of the symmetry groups $G$ 
under the idea that the orbits of $G$ 
are equivalent if they are connected by an isometry of $\M$. 
This leads to the classification of
the conjugacy classes of $G$ in $\Isom \M$. 
The classification is reduced to that of pairs and triples of Killing
vectors which form a Lie algebra. 
We have presented a concrete procedure of the classification. 

We have demonstrated the procedure in Minkowski spacetime 
and have achieved the complete classification of {\co} membranes 
(Tables~\ref{table:result-2dim-comm}, \ref{table:result-2dim-sol} and 
\ref{table:result-3dim}).  
The symmetry group $G$ must be of two or three dimensions in order to
have a two-dimensional orbits. 
In Minkowski spacetime, 
there are two cases for 
$\dim G=2$:
the Abelian group and the non-Abelian group;
and four cases for $\dim G=3$:  
Bianchi type {\VIzero}, {\VIIzero}, {\VIII} and {\IX}. 
The orbits in 
the latter 
four cases are two-dimensional maximally symmetric
timelike or spacelike surfaces.
In addition, because 
$G$ is a subgroup of $\Isom \M$, the embeddings of the 
orbits should be homogeneous and isotropic.
It is interesting that while Bianchi type {\VIzero}, {\VIII} and
{\IX} allow, up to geometric equivalence, 
a unique foliation by the orbits,  
Bianchi type {\VIIzero} allows two inequivalent 
foliations by intrinsically flat orbits. 
One is the flat embedding (Type {\VIIzeroI}) 
and the other is an extrinsically curved one (Type {\VIIzeroII}).  

For the membrane of type {\VIIzeroII},
we have constructed an exact solution. 
The solution describes a $(2+1)$-dimensional analog of 
the flat Friedman-Lema\^itre-Robertson-Walker (FLRW) universe embedded
in Minkowski spacetime.
The cosmological singularity is represented by a null line 
on the {\wv}. 
The embedding is similar to that of the flat
FLRW brane universe in the five-dimensional anti-de Sitter spacetime
\cite{Ishihara:2001qe}.

Our method is general and can be applied to a higher-dimensional
extended object in an arbitrary spacetime.
The equations of motion of an extended object becomes a geodesic
equations on the orbit space $\M/G$. 
As seen in an example of this article, the solution of the geodesic
equations may correspond to a non-trivial configuration of membrane.
Therefore the concept of {\co} objects will be helpful to understand
the dynamics of extended objects in spacetime.

\begin{acknowledgments}
This work is supported by the Grant-in-Aid for Scientific Research
No.19540305 and 24540282. 
\end{acknowledgments}

\appendix*
\section{Equivalence classes of $\Vc^{\mathrm{I}}$}
We give simple representatives of equivalence classes in
$\Vc^{\mathrm{I}}$, which consists of pairs of commuting Killing
vectors $\pair{\bxi_1}{\bxi_2}$, 
where $\bxi_1$ is in the canonical form of Type I, i.e.,
\begin{equation}
  \bxi_1  = a \bP{t} + b \bL{z}, \ \ \ (a, b:\text{arbitrary constant}).
\end{equation}
In the case of $a \neq 0$ and $b\neq 0$,
the general form of $\bxi_2$ which commutes with $\bxi_1$ is simply
\begin{equation}
  \bxi_2 = a' \bP{t} + b' \bL{z} + c' \bP{z}, 
  \label{eq:xi2}
\end{equation}
where $a'$, $b'$ and $c'$ are constants. 
We require 
\begin{align}
 & a b' - b a' \neq 0 \label{eq:case1}
\intertext{or}
 & c' \neq 0 \label{eq:case2}
\end{align}
so that $\bxi_1$ and $\bxi_2$ are linearly independent.
\footnote{In the case of $a = 0$ or $b = 0$,
the general form of $\bxi_2$ is more complicated.
For the sake of simplicity, we concentrate on the case $a, \,b \neq 0$.
}

By virtue of the equivalence under $\GL{2}{R}$ action, 
we can reduce the form of $\pair{\bxi_1}{\bxi_2}$ to a simple one. 
For $\bxi_2$ satisfying Eq.~\eqref{eq:case1},
we have 
\begin{align}
  \pair{\bxi_1}{\bxi_2}
  = \pair{\bP{t} + c \bP{z}}{a' \bP{t} + b' \bL{z} + c' \bP{z}}
  \label{eq:pair_case1}
\end{align}
where $c$ is a constant determined by $a$, $b$, $a'$, $b'$ and
$c'$. 
Otherwise, for $\bxi_2=\bP{z}$, we have
\begin{align}
  \pair{\bxi_1}{\bxi_2} = \pair{a \bP{t} + b \bL{z}}{\bP{z}}.
  \label{eq:pair_case2}
\end{align}

For further reduction of the pair \eqref{eq:pair_case1}, 
we take an isometry $\phi$ generated by $\bK{z}$, namely Lorentz boost
for $z$ direction .
The Killing vector $\bxi_1 = \bP{t} + c \bP{z}$ is transformed to
\begin{align}
  \pf{\phi}\bxi_1
  = (\ch + c \sh) \bP{t} + (c \ch + \sh) \bP{z}.
\end{align}
Choosing the parameter $\theta$ so that 
\begin{align}
  c \ch + \sh &= 0  \ \ \text{for $|c| < 1$}, \\
  \ch + c \sh &= 0  \ \ \text{for $|c| > 1$},
\end{align}
we can reduce the form of the pair \eqref{eq:pair_case1} to
\begin{align}
  \pair{\bxi_1}{\bxi_2}
  = \begin{cases}
    \pair{\bP{t}}{a'' \bP{t} + b' \bL{z} + c'' \bP{z}}
    \ \ \text{for $|c| < 1$, }\\
    \pair{\bP{z}}{a'' \bP{t} + b' \bL{z} + c'' \bP{z}}
    \ \ \text{for $|c| > 1$. }\\
  \end{cases}
\end{align}
In the case $|c| = 1$, $\bxi_1(= \bP{t} \pm \bP{z})$
is invariant under the Lorentz boost $\phi$.
Using $\GL{2}{R}$ action again,
we have three kinds of representatives,
\begin{align}
  \pair{\bxi_1}{\bxi_2}
  = \begin{cases}
    \pair{\bP{t}}{c'' \bP{z} + b' \bL{z}}, \\
    \pair{\bP{z}}{a'' \bP{t} + b' \bL{z}}, \\
    \pair{\bP{t} + \bP{z}}{a'' \bP{t} + b' \bL{z}}.
  \end{cases}
  \label{eq:VCI}
\end{align}
We should note that the pair \eqref{eq:pair_case2}
is included in the second case of \eqref{eq:VCI}.
Therefore these three kinds are the conclusive
representatives of $\Vc^{\mathrm{I}}$ within the case $a,b \neq 0$.
It should also be noted that pairs with different values of constants
$a''$, $b'$ and $c''$ except for overall scaling 
of $\bxi_2$ are not equivalent.


\end{document}